\documentclass{article}
\usepackage{spconf}

		\usepackage[utf8]{inputenc}
		\usepackage{cite}
		\usepackage{graphicx}
		\usepackage{tikz}
			\usetikzlibrary{shapes.geometric}
			\usetikzlibrary{calc}
			\usetikzlibrary{external} 
		\usepackage{tikz-3dplot}
		\usepackage{datatool}
		\usepackage{pgfplots,pgffor}
			\pgfplotsset{compat=newest,every axis/.append style={font=\small},}
		\usepackage{amsmath,amssymb,amsthm}
		\usepackage{url}
		\usepackage{etoolbox}
		\usepackage{xpatch}
		\usepackage{algorithmic}
			\makeatletter
			\xpatchcmd{\algorithmic}{\itemsep\z@}{\itemsep=8pt plus4pt}{}{}
			\makeatother
		\usepackage{array}
		\usepackage[caption=false,font=normalsize,labelfont=sf,textfont=sf]{subfig}
	
		\usepackage{subfiles}
		\usepackage{hyperref}

	\def\natu{\mathbb{N}}
	\def\reals{\mathbb{R}}
	
	\def\sigmax{\sigma_{\max}}
	
	\def\dsigmax{\tilde{\sigma}_{\max}}
	\def\ka{\kappa_\mathrm{a}}
	\def\kd{\kappa_\mathrm{d}}
	\def\pos{\mathbf{r}}
	
	\def\obs{d_{\mathrm{obs}}}
	\def\dobs{\tilde{d}_{\mathrm{obs}}}
	
	\def\dint{\mathrm{d}}
	
	\newcommand{\matrices}[1]{\operatorname{\mathbb{T}}\left( #1 \right)}
	\newcommand{\matricesP}[1]{\operatorname{\mathbb{T}_+}\left( #1 \right)}

	\newcommand{\changetoAlg}{ 
		\renewcommand{\figurename}{Alg.}
	}
	\newcommand{\changetoFig}{
		\renewcommand{\figurename}{Fig.}
	}
	
	\makeatletter%
	\@ifclassloaded{beamer}%
		{}%
		{
		
		}
		\makeatother%

		\def\figs{figs}
		
		\def\bib{.}
	
\title{Cell Detection on Image-based Immunoassays}
\name{Pol del Aguila Pla\thanks{Thanks to the KTH Opportunities Fund for travel funding, and to Mabtech AB for funding and data. Thanks to the Swedish Research Council (VR) for funding (grant 2015-04026).} and Joakim Jaldén}
\address{School of Electrical Engineering,
		 Department of Information Science and Engineering \\
		 KTH Royal Institute of Technology, Stockholm, Sweden \\
		 \href{mailto:poldap@kth.se}{\nolinkurl{[poldap,jalden]@kth.se}} }

		\def\figs{figs}
		\def\bib{.}		

\begin{document}
%
\maketitle
\begin{abstract}
		Cell detection and counting in the image-based ELISPOT and Fluorospot 
	immunoassays is considered a bottleneck.
	The task has remained hard to automatize, 
	and biomedical researchers often have to rely on results that are not accurate. 
	Previously proposed solutions are heuristic, and data-based
	solutions are subject to a lack of objective ground truth data. 
	In this paper, we analyze a partial differential equations model
	for ELISPOT, Fluorospot, and assays of similar design. 
	This leads us to a mathematical observation model for the images generated by 
	these assays. We use this model to motivate a methodology for cell
	detection. Finally, we provide a real-data example that suggests that this cell detection 
	methodology and a human expert perform comparably.

\end{abstract}
\section{Introduction}
\label{sec:intro}

	In this paper, we use a well-known physical partial differential equations (PDE) model \cite{Lagerholm1998,Lieto2003,Berezhkovskii2004,Plante2011,Karulin2012}
	to obtain an observation model \cite{AguilaPla2017,AguilaPla2017a} that contributes to the analysis and synthesis of 
	data from image-based immunoassays such as ELISPOT \cite{Czerkinsky1983} and Fluorospot \cite{Gazagne2003}. These immunoassays are relevant to 
	pharmacological development and medical research \cite{Dillenbeck2014,Martinez-Murillo2016}, and can even be used to diagnose certain diseases 
	\cite{Meier2005,2015}.
	
	The data that result from the considered immunoassays are noisy images containing spots of different shapes and sizes, which may overlap 
	and occlude each other (see Fig.~\ref{fig:physics} for an example section). From a biological perspective, the most relevant information
	in these images is the number of spots they contain and their precise location. The former is used to establish which proportion of the cells involved in 
	an experiment secreted a substance of interest, while the latter is used to correlate this information with parallel assays for some other substance on the same 
	cell population (multiplex assays). For example, in \cite{Dillenbeck2014}, a Fluorospot assay was run for the cytokines 
	IFN-$\gamma$, IL-$17$A and IL-$22$ to determine the proportion of human peripheral blood mononuclear cells that generated one, two or all of these
	substances under the effect of a specific antigen.
	
	In conclusion, accurate detection and localization of the spots in these images is critical to the validity of the results and conclusions extracted from 
	these assays, even more so in the case of multiplex assays. However, approaches to spot detection generally rely on heuristic methods to find dot-like
	shapes combined with generic methodologies to address measurement noise \cite{Rebhahn2008,Smal2010}. In this paper, we use the aforementioned PDE model
	to obtain an observation model for the resulting images, and, through it, a well-founded methodology for cell detection.
	
\section{From PDE to imaging}

\begin{figure*}
		\centering
		\begin{tabular}{cc}
			\includegraphics[keepaspectratio=true,clip=true,trim=3in 8.97in 3.4in 1.04in,height=1.4625in]{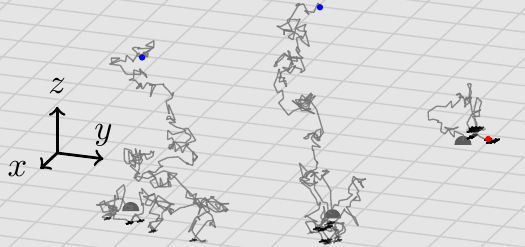} & \includegraphics[keepaspectratio=true,height=1.4625in]{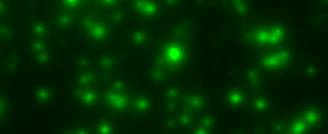} \\
			\small (a) Particles' motion model  & \small (b) Typical observation \\
		\end{tabular} \vspace{-12pt}
		\caption{ \small
				(a) Visualization, at a particle level, of the proposed physical data model.  
				Three particles, each secreted by a different cell (dark gray) immobilized on the plane (light gray), follow a Brownian motion.
				When they hit the plane, they might bind to it (adsorption; black marks). After a time, they may disassociate
				(desorption) and continue their Brownian motion. At the end of the experiment, i.e., at time $T$, they may be free (blue dots) and thus 
				not imaged, or bound to the surface (red dot) and thus contribute to the final image. 
				(b) Example section of an image observation from a Fluorospot assay. 
				FITC dye, marking IFN-$\gamma$ molecules, emitting $512~\mathrm{nm}$ fluorescence. 
				Image captured by an RGB camera.  \vspace{-10pt}
				\label{fig:physics} 
				} 
	\end{figure*}

	The spots in the considered images are the result of particles generated by cells (reaction) during a time window $[0,T]$. These cells (hereon, active cells) are immobilized 
	at the bottom of a well, i.e. on the plane $z=0$. The particles they generate undergo a Brownian motion through a medium (diffusion), modeled here by the half-space $z\geq 0$. 
	When these particles collide with the plane $z=0$, they can bind to an even coat of receptors that covers it (adsorption),
	and after some time, they can break the bond and continue their motion (desorption). At time $T$, the experiment finishes and the density of bound particles is imaged. Fig.~\ref{fig:physics}
	exemplifies this physical model at a particle level and exhibits a section of a real image from a Fluorospot immunoassay.
		
	From a macroscopic point of view, this model can be expressed in terms of the time-varying concentration of particles that move freely 
	in $z>0$, i.e., $c(x,y,z,t)\geq0~[\mathrm{m}^{-3}]$. This density follows the diffusion equation 
	\begin{subequations} \label{eq:pde}
		\begin{equation} \label{eq:pde-diffusion}
			\frac{\partial}{\partial t} c = D \left( \frac{\partial^2 c}{\partial x^2} + \frac{\partial^2  c}{\partial y^2} +  \frac{\partial^2 c}{\partial z^2}\right) 
			\,, \\
		\end{equation}
		subject to boundary conditions at $z=0$ that express reaction, adsorption and desorption. These boundary conditions couple $c(x,y,z,t)$ to 
		the surface density of bound particles at time $t$, i.e., $d(x,y,t)\geq0~[\mathrm{m}^{-2}]$, and the source density 
		rate (SDR) of new particles generated by cells residing at surface locations, i.e., $s(x,y,t)\geq0~[\mathrm{m}^{-2} \mathrm{s}^{-1}]$, 
		through \cite{Karulin2012}
		\begin{equation}
		 \label{eq:pde-surface}
			\frac{\partial}{\partial t} d = \kappa_\mathrm{a} c \big|_{z=0} - \kappa_\mathrm{d} d\,,
		\end{equation}
		and
		\begin{equation}
			\label{eq:pde-boundaryflow}
			-D \frac{\partial}{\partial z} c \big|_{z=0} = s + \kappa_\mathrm{d} d - \kappa_\mathrm{a} c\big|_{z=0}\,.
		\end{equation}
	\end{subequations}
	Here, $\ka~[\mathrm{m}\mathrm{s}^{-1}]$, $\kd~[\mathrm{s}^{-1}]$, and $D~[\mathrm{m}^{2}\mathrm{s}^{-1}]$ are physical parameters
	characterizing the surface's adsorption and desorption rates and the medium's diffusion constant, respectively.
	More on the generality and assumptions of this model can be found in \cite{Lagerholm1998,Lieto2003,Berezhkovskii2004,Plante2011,Karulin2012,AguilaPla2017}. 
	Note that $s(x,y,t)$ is spatially sharp and sparse, because particles are only released from locations occupied by cells, but also temporally continuous, because cells are immobilized throughout the experiment.
	
	In \cite{AguilaPla2017}, we prove that the luminosity function of the captured image, i.e., the observation $\obs(x,y)$, can be expressed up 
	to a constant of proportionality as
	\begin{equation} \label{eq:observation-model}
		\obs(x,y) = d(x,y,T) = \int_0^{\sigmax}\hspace{-10pt} g_{\sigma}(x,y) * a(x,y,\sigma) \dint\sigma ,
	\end{equation}
	where $g_\sigma$ is a 2D isotropic Gaussian kernel of standard variation $\sigma$, $\sigmax = \sqrt{2DT}$ and $*$ represents spatial convolution.
	$a(x,y,\sigma)\geq 0$ for $\sigma\geq 0$ is a new quantity that we name post adsorption-desorption source density rate (PSDR), and that can be expressed as 
	\begin{equation} \label{eq:weights-a}
		a(x,y,\sigma) = \frac{\sigma}{D} \int_{\frac{\sigma^2}{2D}}^T s(x,y,T-\eta) \, \varphi\!\left(\frac{\sigma^2}{2D},\eta\right) \dint\eta\,.
	\end{equation}
	The PSDR expresses an equivalent SDR where the effect of adsorption and desorption has been summarized. Moreover, the PSDR is expressed
	as a function of the length $\sigma$ each of the particles has traveled from the site they were released, as opposed to the SDR, which is
	a function of the time at which each particle was released.
	In \cite{AguilaPla2017a}, we prove that $ \varphi(\tau,t)$ in \eqref{eq:weights-a} is given by
	\begin{equation} \label{eq:varphi}
		\varphi(\tau,t) = i_{[0,t)}(\tau) \sum_{j=1}^\infty \phi^{j*}(\tau) p\left[j-1;\kd(t-\tau)\right]\,,
	\end{equation}
	for $0\leq \tau \leq t$, $\forall t \leq T$. This function expresses the probabilistic relation between the total time in free motion 
	$\tau$ and the time $t$ at which a particle is found bound. In \eqref{eq:varphi}, we have that $p[j;\lambda]$ is the probability
	mass function of a Poisson random variable with mean $\lambda\geq 0$ evaluated at $j\in\natu$, $i_{[0,t)}(\tau)$ is the $(0,1)$-indicator function of the set $[0,t)$,
	\begin{equation*}\label{eq:first-adsorption}
			\phi(\tau) =  \frac{\ka}{\sqrt{\pi D \tau}} 
			- \frac{\ka^2}{D} \mathrm{erfcx}\left( \ka \sqrt{\frac{\tau}{D}} \right) \,, 
	\end{equation*}
	and $\phi^{j*}(\tau) = (  \overbrace{\phi *\cdots * \phi}^j )(\tau)$ is $\phi$'s $j$-th convolutional product. 
	Here, $\mathrm{erfcx}(x)$ is the scaled-complementary error function. For more on the generality of this observation model and 
	how it is affected by hardware impairments such as optical blur or additive noise, see \cite{AguilaPla2017a}.
	
	For the purpose of cell detection, it is relevant to note that $a(x,y,\sigma)$ contains 
	the same spatial information that $s(x,y,t)$ does, because
	the operation to obtain $a(x,y,\sigma)$ from $s(x.y,t)$ \eqref{eq:weights-a} is only a convolution
	in the temporal dimension, which leaves spatial dependence unchanged.
	Consequently, $a(x,y,\sigma)$ is also spatially sharp and sparse, indicating where active
	cells lie. Inverting \eqref{eq:observation-model} to obtain the PSDR, then, is a
	reasonable procedure for cell detection. Moreover, recovering the PSDR also provides a 
	representation of the amount of particles released from each cell location.
	For the purpose of understanding spot formation, one can simply picture the response of the observation model \eqref{eq:observation-model} to a spatially sharp, temporally continuous
	$s(x,y,t)$. This reveals that spots that are generated by active cells will always be monotone and circularly invariant.
	For the purpose of synthetic data generation, note that \eqref{eq:observation-model}, \eqref{eq:weights-a} and \eqref{eq:varphi} provide all the information needed to generate a synthetic image
	observation from an arbitrary SDR $s(x,y,t)$ and some physical parameters $\ka,\kd,D$ and $T$.
	
	In conclusion, the novel observation model \eqref{eq:observation-model} allows for a new understanding of the 
	spot formation process, but also provides natural analysis and synthesis strategies.

\section{Methodology}
\label{sec:met}

	\subsection{Inverse Problem}
	
		In \cite{AguilaPla2017}, we derive a procedure for inverting the observation
		model \eqref{eq:observation-model}
		in function spaces by using group-sparsity regularization and the 
		accelerated proximal gradient (APG) algorithm (also known as FISTA).
		In \cite{AguilaPla2017a}, we propose a discretization and approximation
		of that algorithm. This discretized algorithm
		obtains a discrete approximation $\tilde{a}\in\matricesP{M,N,K}$ of 
		$a(x,y,\sigma)\geq 0$ in \eqref{eq:weights-a} from a discrete image
		observation $\dobs\in\matricesP{M,N}$. Here, $\matricesP{q_1,q_2,\dots,q_Q}$ is the space of element-wise non-negative tensors (or matrices) of dimension 
		$q_1 \times q_2 \times \dots \times q_Q$, $M$ and $N$ are the number of 
		pixels in each spatial direction, and $K$ is
		the number of discretization points used for the $\sigma$-dimension.
				
		\begin{figure} \small
			\begin{algorithmic}[1]
			\vspace{.5pt}\hrule height 1pt \vspace{.5pt}
			\REQUIRE { \small An initial $\tilde{a}^{(0)}\in\matrices{M,N,K}$, a discrete image observation $\dobs\in\matrices{M,N}$}
			\vspace{.5pt}\hrule height .5pt \vspace{.5pt}
				\item[] 
				\STATE $\tilde{b}^{(0)} \leftarrow \tilde{a}^{(0)}$, $i\leftarrow 0$ \label{line:iniacc}
				\REPEAT \vspace{3pt}
					\STATE $i\leftarrow i+1$ \label{line:acc1} \vspace{-8pt}
					\STATE $\displaystyle \tilde{d}^{(i)} \leftarrow \sum_{k=1}^{K} \tilde{g}_k \circledast \tilde{b}^{(i-1)}_k - \dobs$ \label{line:forwardanddiff} \vspace{-5pt}
					\FOR{ $ k=1 $ \TO $K$}  \vspace{6pt}
						\STATE $\displaystyle \tilde{a}_k^{(i)} \leftarrow \left[ \tilde{b}_k^{(i-1)} - \eta \tilde{g}_k \circledast \left[\tilde{w}^2 \odot \tilde{d}^{(i)}\right] \right]_+$ \label{line:adjointandpos}
						\vspace{-8pt}
					\ENDFOR
					\STATE $\displaystyle \tilde{p} \leftarrow \left( 1 - \frac{\eta}{2} \lambda \left[ \sqrt{\sum_{k=1}^{K} \left(\tilde{a}^{(i)}_{k}\right)^2}\right]^{-1} \right)_+ $ \label{line:shrinkth1}
					\FOR{ $ k=1 $ \TO $K$} \vspace{3pt}
						\STATE  $\displaystyle \tilde{a}_k^{(i)} \leftarrow \tilde{p} \odot \tilde{a}_k^{(i)} $ \label{line:shrinkth2}
						\vspace{-5pt}
					\ENDFOR
					\STATE $\tilde{b}^{(i)} \leftarrow \tilde{a}^{(i)} + \alpha(i) \left( \tilde{a}^{(i)} - \tilde{a}^{(i-1)}\right)$ \label{line:acc2}
				\UNTIL{ convergence }
				\STATE $ \tilde{a}_{\mathrm{opt}} \leftarrow \tilde{a}^{(i)}$
				\item[]
				
			\vspace{1pt}\hrule height 1pt \vspace{-10pt}
			\end{algorithmic}
			\caption{ \small APG algorithm to obtain $\tilde{a}$. 
					Lines~\ref{line:forwardanddiff} and \ref{line:adjointandpos} optimize the data fidelity term, 
					while Lines~\ref{line:shrinkth1} and \ref{line:shrinkth2} optimize the regularizer. 
					The sequence $\alpha(i)$ can be that in \cite{Beck2009} or that in \cite{Chambolle2015}, $\eta = \dsigmax^{-1}/\max|\tilde{w}_{m,n}|^{2}$ is the algorithm's fixed step size, 
					$\circledast$ represents discrete size-preserving zero-padded convolution, and matrix products ($\odot$) and powers are element-wise.
					\label{algs:AccProxGradforRegInvDif}  \vspace{-10pt}
					} 
		\end{figure}
				
		Fig.~\ref{algs:AccProxGradforRegInvDif} specifies a simplified case of this discrete algorithm. 
		The $\tilde{g}_k$s are discrete rank-$1$ convolutional kernels formed by approximating finite integrals of Gaussian functions with respect to their standard deviation
		and spatial coordinates (see $g_k^{\mathrm{b1r}}$ in \cite{AguilaPla2017a} for details), and the $\tilde{a}_k$s are cuts of $\tilde{a}$ in the $k$-dimension, i.e.
		$\tilde{a}_k\in\matricesP{M,N}$. Moreover, $\dsigmax=\sigmax / \Delta_{\mathrm{pix}}$, where $\Delta_{\mathrm{pix}}$ is the length
		of a pixel's side, and $\tilde{w}\in\matricesP{M,N}$ and $\lambda\geq 0$ are user parameters. In particular, 
		the algorithm in Fig.~\ref{algs:AccProxGradforRegInvDif} solves the finite-dimensional optimization problem
		\begin{equation} \label{eq:optidisc}
			\min_{\tilde{a}}
			\left\lbrace 
				\left\| \tilde{w} \odot\! \left( \dobs\! -\! \sum_{k=1}^{K} \tilde{g}_k \circledast \tilde{a}_k \right) \right\|_2^2
				\! + \! \lambda \sum_{m,n} \! \left\| \tilde{a}_{m,n} \right\|_2
			\right\rbrace
		\end{equation}
		subject to $\tilde{a}\in\matricesP{M,N,K}$, where the $\tilde{a}_{m,n}$s are cuts of $\tilde{a}$ in the spatial dimensions, i.e. $\tilde{a}_{m,n}\in\reals^K$.
		This optimization problem can be proven to approximate the one proposed in \cite{AguilaPla2017}.
		The first term in \eqref{eq:optidisc} is a weighted norm used as a data-fidelity
		cost function with respect to a discretization of the observation 
		model \eqref{eq:observation-model}. The second term in \eqref{eq:optidisc} is
		a regularizer that promotes both spatial sparsity and continuity through the $k$s, i.e.,
		a group-sparsity regularizer that induces a group behavior \cite{Yuan2006}
		for all the components in $\tilde{a}$ representing a certain location.
		The effect of this regularizer is tweaked by the regularization parameter
		$\lambda$, which is set larger (or lower) to increase (or decrease) selectivity.

	\subsection{Detection and performance evaluation}
	
		In the context of image-based immunoassays, a cell detector generally provides tuples 
		$\left\lbrace (\pos_l, p_l) \right\rbrace_{l=1}^{L}$, where $\pos_l\in\reals_+^2$ is
		a position in pixel-based coordinates and $p_l\geq 0$ is a non-negative number proportional to the
		confidence assigned to the specific detection, i.e. a pseudo-likelihood. 
		This is done so that researchers can threshold detections by the pseudo-likelihood to match their 
		criteria.
		
		In our specific case, we use the estimated discrete PSDR $\tilde{a}$ obtained from 
		the algorithm in Fig.~\ref{algs:AccProxGradforRegInvDif} to build an image 
		$\tilde{p}=\sqrt{\sum_{k} \tilde{a}^{2}_{k}}$. Then, we consider the pixel positions of its regional 
		maxima as detections, and the pixel values at those positions as the respective pseudo-likelihoods.
		This specific $\tilde{p}$ expresses the importance of each detection with respect to the group-sparsity 
		regularizer in \eqref{eq:optidisc}.
		
		\def\TP{\mathrm{TP}} \def\FP{\mathrm{FP}} \def\FN{\mathrm{FN}}
				\begin{figure*}
		\centering
		\includegraphics[width=.495\linewidth,keepaspectratio=true]{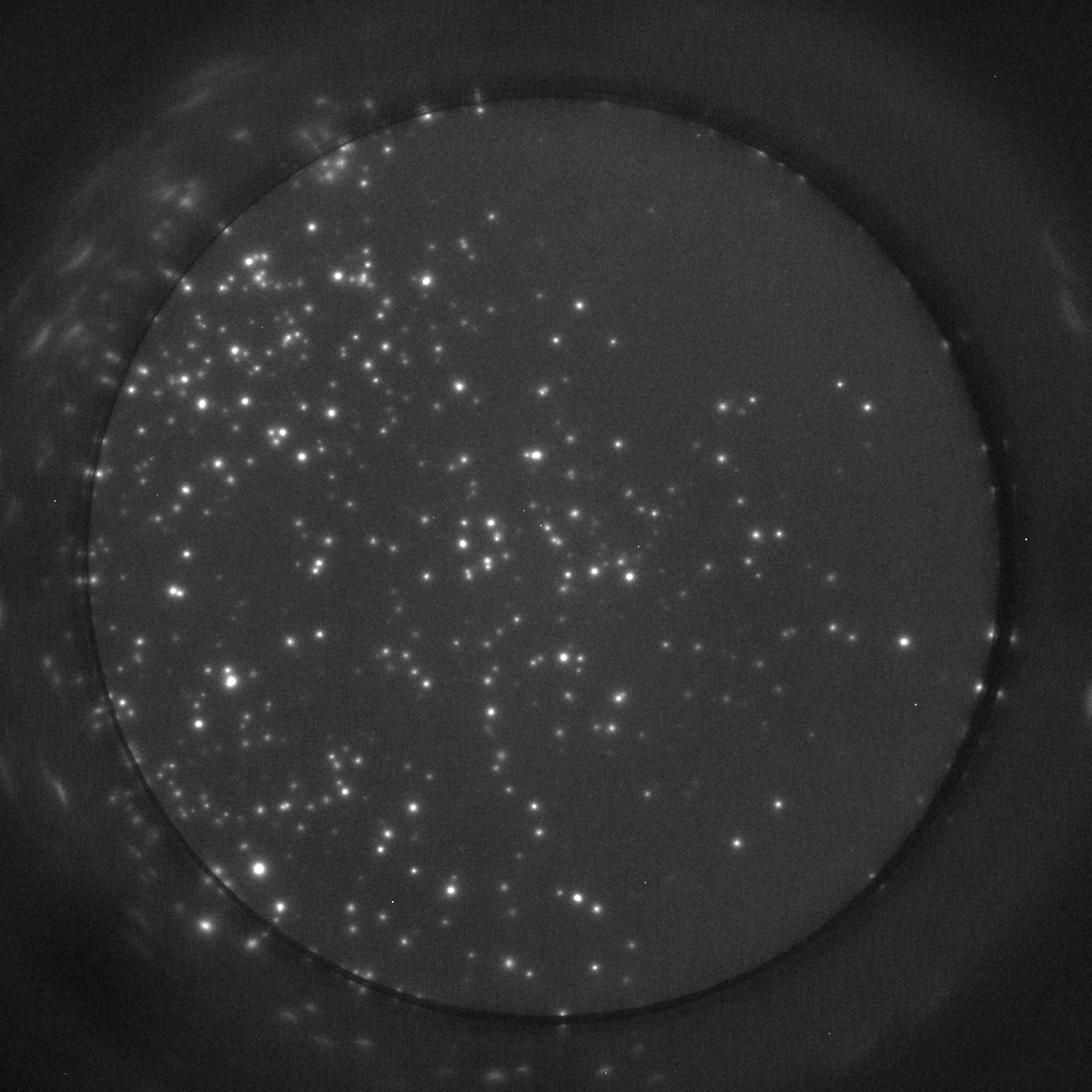} 
		\includegraphics[width=.495\linewidth,keepaspectratio=true]{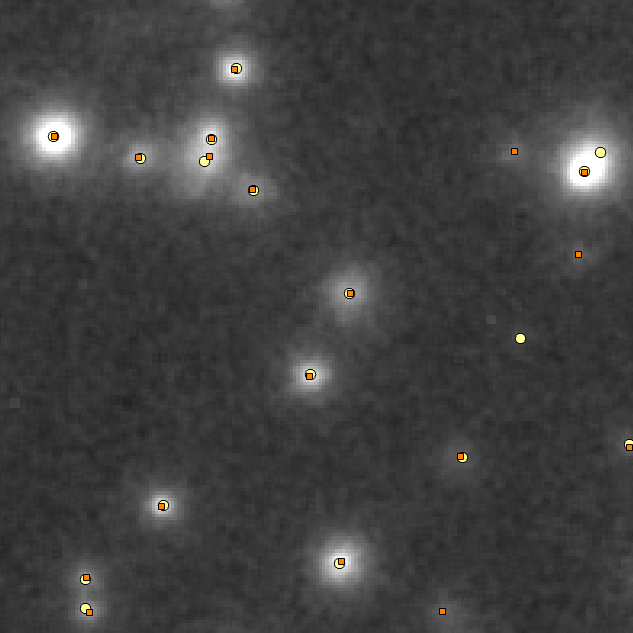}
		\caption{ \label{fig:dataexample}
			To the left, grayscale image representation of the data, with increased 
			luminosity. To the right,  cell detection results (yellow circles) and human expert labeling 
			(orange squares) for a specific section. 
		}
	\end{figure*}
			
		To evaluate the performance of our detector when ground-truth data is available, we pick the threshold for 
		the pseudo-likelihoods $p_l$ that yields the best F1-score given the ground-truth data. In this manner, we hope to 
		emulate the threshold the human expert would have chosen. The F1-score is
		a number in the range $[0,1]$ that expresses a compromise between precision and recall, i.e., 
		\begin{equation*}
				\mathrm{pre} = \frac{\TP}{\TP + \FP},\, \mathrm{rec} = \frac{\TP}{\TP + \FN},\mbox{ and }
				\mathrm{F1} = \frac{2 \, \mathrm{pre} \cdot \mathrm{rec}}{\mathrm{pre}+ \mathrm{rec}},\,
		\end{equation*}
		with $\TP$, $\FP$ and $\FN$ the numbers of true and false positives and false negatives, respectively. These 
		quantities are obtained by matching the detections to ground-truth cell positions in decreasing
		order of pseudo-likelihood with a tolerance of $3~\mathrm{pixels}$.

\section{Real-Data Example}
\label{sec:exp}

		We analyzed a real Fluorospot image for which human expert labeling was available. This image was obtained by using FITC dye as a marker for some 
	relevant analyte, and was captured by an RGB sensor that yielded a $2048\times 2048$ raw image with a dynamic range of $16~\mathrm{bits}$. 
	The data was subject to a Bayer filter, i.e., neighboring pixels exhibited different sensitivities to light intensity at the 
	FITC wavelength ($512~\mathrm{nm}$). To compensate this difference, we weighted each pixel correspondingly to estimate the luminosity, 
	and used $\tilde{w}$ to weight the prediction error at each pixel according to its sensitivity. 
	Furthermore, we selected the area that comprised the well manually, and fixed $\tilde{w}=0$ for all points outside it. 
	
	We used the algorithm in Fig.~\ref{algs:AccProxGradforRegInvDif} with $M=N=2048$, $K=6$, $\lambda=4000$ and the sequence $\alpha(i)$
	proposed in \cite{Beck2009}. The underlying parameters $\sigma_k$ (see \cite{AguilaPla2017a}) were set to 
	$\lbrace 2, 15, 20, 30, 40, 50, 70 \rbrace$. We run the algorithm for 10000 iterations, which were more than those needed
	for convergence. The resulting F1-Score was $0.9$ with precision $0.92$ and recall $0.88$. 
	On the left panel of Figure~\ref{fig:dataexample}, we show a grayscale representation of the image under study, while on the right panel,
	we show both the detections proposed by the human expert (orange squares) and the ones proposed by 
	our algorithm (yellow circles), on a specific section of the image.
	
	In our opinion, both sets of detections are of comparable quality, with 
	our algorithm being more precise in terms of cell locations and the human labeler obtaining higher recall
	for isolated cells. However, one has to take into account that the detections obtained by our algorithm 
	have been thresholded to match the criteria of this specific expert, and thus, the absence of weaker spots in the set of detections
	can be explained by inconsistent inclusion criteria in the human labeling. 
	A final relevant difference between the two 
	sets of detections is that our algorithm uses the observation model to 
	evaluate the whole shape of spots in
	terms of possible cells, instead of mainly relying on local luminosity. 
	Hence, the algorithm includes detections that are weaker
	but fit the shape of cell-generated spots, as the apparent false positive
	in the middle-right region of the image.
	This also results in the correct decomposition of clusters of cells, as 
	it is clearly the case of the large spot in the upper-right region of 
	the image. 
	
	The results reported here are coherent with the extensive quantitative study on synthetic data
	we present in \cite{AguilaPla2017a}, which additionally suggests robustness both to additive noise and to changes in the
	regularization parameter $\lambda$, as well as dominance over simpler deconvolution approaches.

\section{Conclusions}
\label{sec:con}

	In this paper, we have analyzed the PDE behind some image-based 
	immunoassays, i.e. a reaction-diffusion-adsorption-desorption equation. 
	From this analysis, we have obtained a novel observation model for these 
	assays. Then, we have presented the insights on the process of 
	spot formation this observation model entails, and we have used them
	to propose a novel analysis algorithm. 
	Finally, we have 
	exemplified
	the use of this algorithm on real data, obtaining results that are 
	quantitatively close and qualitatively comparable to those generated 
	by a human expert.

\bibliographystyle{IEEEbib}

{\small
\bibliography{\bib/multi_deconv}}

\end{document}